\documentclass[cameraready]{Interspeech}

\title{Mitigating Speaker Leakage in Cascaded Multi-talker ASR with Diarization-based Transcript Correction}

\author[affiliation={1}, orcid=0000-0003-2861-4891]{Hermann}{Yepdjio Nkouanga}
\author[affiliation={1}, orcid=0009-0005-2337-7163]{Minwei}{Luo}
\author[affiliation={2}, orcid=0000-0003-1707-8106]{Maggie}{Wigness}
\author[affiliation={1}, orcid=0000-0002-1509-0497]{Suresh}{Singh}

\address{
    $^1$ Portland State University, USA \\
    $^2$ US Army Research Laboratory, USA 
}

\email{hermann@pdx.edu, minwei@pdx.edu, maggie.b.wigness.civ@army.mil, singh@cs.pdx.edu}


\usepackage{comment}
\usepackage{multirow, xcolor, graphicx, booktabs, amssymb}

\begin{document}

\maketitle

\begin{abstract}
While cascaded multi-talker ASR (MT-ASR) leverages state-of-the-art foundation models, its performance is often capped by speaker leakage during separation. Prior correction strategies primarily focus on lexical re-labeling for speaker attribution. We propose a complementary pruning-based paradigm that robustly identifies and removes leakage artifacts. Our method utilizes a pre-trained speaker diarization model as a multimodal verifier to prune transcribed segments satisfying a tripartite consensus of temporal containment, lexical cross-validation, and temporal alignment. Results on LibriMix, LibriSpeechMix, and the AMI Meeting corpus show our algorithm consistently reduces $cpWER$ across diverse overlap conditions. Specifically, on subsets with high speaker leakage, our method achieves relative $cpWER$ reductions of up to 29\%, highlighting its effectiveness in enhancing the reliability of cascaded MT-ASR transcripts in complex acoustic environments.
\end{abstract}

\section{Introduction}
\label{introduction}

Automatic Speech Recognition (ASR), the process of converting spoken words into text, has achieved high accuracy and is finding increasing use in daily life. However, a variant of ASR that consists of transcribing overlapping speeches from different speakers remains an open research problem. It is often referred to as multi-talker ASR (MT-ASR) or the cocktail party problem. Numerous solutions to this variant have been proposed, which often fall under one of two main categories: end-to-end (E2E) and cascaded. In E2E systems \cite{Meng2024whisper, Thai-Binh2024, Wang2025SpeakerTV, shi2025serializedoutputprompting}, a single system is trained to map an audio mixture directly to a multi-talker transcript. These systems mitigate error propagation as their components are trained jointly. On the other hand, the cascaded method \cite{Yufeng2025, Paturi2021DirectedSS, continuous_ss_w_conformer2021} typically involves separating the speech sources first, followed by transcription using an ASR model. The main advantage of this method lies in its modularity, which provides the ability to leverage foundation models while remaining adaptable for various downstream tasks. 

A major limitation of cascaded systems is that their performance is fundamentally capped by the quality of the initial separation. Imperfect separation inevitably leads to artifacts being left in the separated sources. A common outcome of this imperfection is ``leakage" \cite{vieting2025error}, where residual speech from interfering speakers contaminates the output streams, causing significant errors in the final speaker attribution.

To address the challenge of speaker leakage, several post-processing strategies have been explored in the literature. The dominant paradigm focuses on re-labeling misattributed words. Early methods were purely lexical and relied only on text \cite{Paturi2023LexicalSE}. They have since evolved to include powerful but computationally expensive Large Language Models (LLMs) that ``proofread" transcripts \cite{Wang2024DiarizationLMSD}. The most advanced methods create a hybrid, fusing textual context with acoustic information to make more robust re-labeling decisions \cite{paturi2024aglsec, Kumar2025SEALSE}.

In this paper, we diverge from these re-labeling strategies and instead introduce and evaluate a complementary pruning-based approach. Our method focuses on robustly identifying and removing leakage artifacts altogether. We propose a novel correction algorithm that employs an independent, pre-trained speaker diarization model as a verifier on each separated audio stream. Any words transcribed within segments identified as containing an interfering speaker are then pruned from the final transcript. Furthermore, using insights from this algorithm, we investigate an exploratory model architecture that incorporates a diarization head into an existing speech separation model to assess whether speaker leakage can be mitigated during the separation stage. While our experiments reveal that the joint architecture remains sensitive to domain shifts, the proposed pruning algorithm consistently reduces the Concatenated minimum-Permutation Word Error Rate ($cpWER$) by up to 29\% under high-leakage conditions.

Our contributions can be summarized as follows:
\begin{itemize}
\item We propose a novel, diarization-based pruning algorithm to detect and correct speaker leakage errors in cascaded MT-ASR systems.
\item We demonstrate the effectiveness of our proposed method through experiments, showing a significant reduction in the $cpWER$ on various multi-talker datasets.
\item We evaluate an experimental multi-task architecture incorporating a diarization head into a separation backbone, analyzing its potential to suppress leakage at the signal level and its generalization challenges.

\end{itemize}

\section{Related Work}
\label{related_work}

The challenge of accurately transcribing multi-talker audio has been approached from multiple angles. Our work, which focuses on correcting errors in cascaded ASR systems, is situated within a broad landscape of architectural paradigms and post-processing techniques.

\subsection{Architectural Paradigms in Multi-Talker ASR}
MT-ASR systems generally follow end-to-end (E2E) or cascaded designs. E2E models \cite{Meng2024whisper, Thai-Binh2024, Wang2025SpeakerTV, shi2025serializedoutputprompting} jointly optimize recognition and diarization to produce a single stream with speaker markers, reducing error propagation but requiring massive multi-speaker datasets. Conversely, cascaded systems \cite{Yufeng2025, Paturi2021DirectedSS, continuous_ss_w_conformer2021} chain independent separation and ASR models. While modular and adaptable to foundation models, these systems are fundamentally limited by separation quality from the first stage of the pipeline.

\subsection{Speaker Error Correction}
Given that cascaded systems can suffer from speaker leakage, a significant amount of research has focused on post-processing their output to correct speaker attribution errors. Early lexical methods (LSEC) \cite{Paturi2023LexicalSE} used text context and first-pass diarization for word-level classification, though their lack of acoustic grounding often led to over-correction. Recently, LLMs have been explored. While some research used them to re-diarize transcripts based on semantic logic and Chain-of-Thought reasoning \cite{Wang2024DiarizationLMSD, Adedeji2024TheSoundOfHealcare}, others integrated their probabilities directly into the decoding process via contextual beam search, fusing acoustic and lexical scores in real-time\cite{Park2023EnhancingSD}. However, an issue with LLM-based correction is that they introduce latency and hallucination risks.

To improve the robustness of text-only methods, hybrid methods have also been explored to incorporate acoustic evidence. For instance, AG-LSEC \cite{paturi2024aglsec} fuses lexical embeddings with frame-level posterior scores, while SEAL \cite{Kumar2025SEALSE} prompts LLMs with discretized acoustic confidence scores. 

Beyond re-labeling, alternative strategies include ``Constrained Diarization" \cite{Turn-to-Diarize2022}, which uses speaker turn tokens as guidance, and human-in-the-loop interfaces \cite{He2025InteractiveRS} for manual correction.

\subsection{Positioning of the Proposed Work}
While existing literature is dominated by the re-labeling paradigm, our work evaluates a pruning-based strategy. We hypothesize that for significant speaker leakage, re-labeling is inherently ambiguous. Our approach utilizes an independent diarization model to verify and prune corrupted segments, offering a direct, efficient solution for transcript integrity. Additionally, we evaluate a preventative framework incorporating a diarization head into a separation backbone to investigate signal-level mitigation.

\section{Methods}
\label{methods}

In this section, we present our proposed multimodal diarization-based pruning algorithm for speaker leakage correction. We also provide details for our exploratory multi-task architecture designed for signal-level correction.

\subsection{Diarization-based Transcript Correction}
\label{diarization_based_correction_method}


Our correction algorithm is formalized as follows: Let $A$ and $B$ represent the separated audio streams for Speaker $1$ and Speaker $2$, respectively. Let $\mathcal{T}_1$ and $\mathcal{T}_2$ be the initial transcripts generated by the ASR model, where each transcript is a set of tuples $\mathcal{T} = \{ (w_i, t_{s,i}, t_{e,i}) \}$, representing the word, start time, and end time.

\textbf{Acoustic Leakage Detection:} diarization is applied to $A$ and $B$ to identify segments of interfering speech. For stream $A$, the diarizer produces a set of leakage windows $\mathcal{W}_A = \{ [T_s, T_e] \}$, representing intervals where the model detects the presence of the interfering speaker.

\textbf{Correction Logic:} A tuple $(w_i, t_{s,i}, t_{e,i}) \in T_1$ is flagged as leakage if it satisfies a \textit{tripartite consensus} of acoustic, lexical, and temporal evidence:

\begin{itemize}
    \item \textbf{Acoustic Containment ($C_{\text{ac}}$):} The word is temporally contained within a detected leakage window $W \in \mathcal{W}_A$.
    \item \textbf{Lexical Cross-Validation ($C_{\text{lex}}$):} The word $w_i$ exists in the parallel transcript $T_2$.
    \item \textbf{Temporal Alignment ($C_{\text{temp}}$):} There exists a tuple $(w_j, t_{s,j}, t_{e,j}) \in T_2$ such that $w_i = w_j$ and the intervals overlap:
    \begin{equation}
        \begin{split}
            C_{\text{temp}} : & \exists (w_j, t_{s,j}, t_{e,j}) \in T_2 \text{ s.t. } (w_i = w_j) \\
            & \land (t_{s,i} \le t_{e,j} \land t_{e,i} \ge t_{s,j})
        \end{split}
    \end{equation}
\end{itemize}

The corrected transcript $T'_1$ is then formed by pruning only the words where all three conditions hold true:
\begin{equation}
    T'_1 = \{ (w, t_s, t_e) \in T_1 \mid \neg (C_{\text{ac}} \land C_{\text{lex}} \land C_{\text{temp}}) \}
\end{equation}

\begin{figure}[h!]
  \centering
  \includegraphics[width=\linewidth]{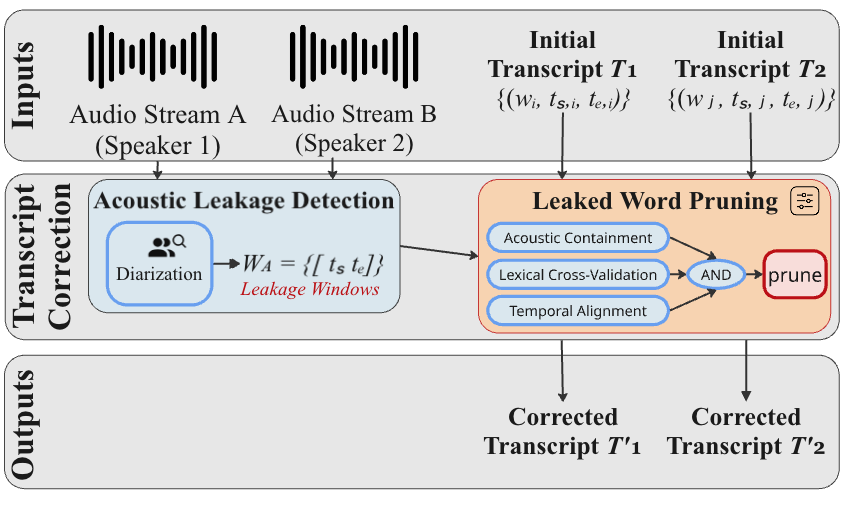}
  \caption{Overview of the proposed transcript correction pipeline. The system identifies acoustic leakage using diarization, followed by a tripartite consensus check to prune leaked words.}
  \label{fig:system_diagram}
\end{figure}

The overall workflow of this correction algorithm is summarized in Figure \ref{fig:system_diagram} 

\textbf{Adaptive Filtering:} To ensure robustness, the correction is only applied if the global similarity $\sigma(\mathcal{T}_1, \mathcal{T}_2)$ between transcripts exceeds a threshold $\gamma$. We define this similarity using the Ratcliff/Obershelp pattern recognition algorithm \cite{black2004ratcliff}. If $\sigma < \gamma$, the separation is deemed successful and correction is bypassed. In our experiments, we set $\gamma = 0.40$ as it provided the optimal performance after evaluating a range of values between $0.30$ and $1.0$.

\subsection{Joint Mossformer-Diarization Framework}
\label{joint_mossfomer_diarization_method}
We also evaluate an experimental multi-task framework that augments a separation model---MossFormer2 \cite{Mossformer2024}---with a diarization head operating on shared encoded features. The encoder generates 512-dimensional frame-level representations via 24 interleaved iterations of FLASH self-attention and Gated Feedforward Sequential Memory Networks. These features are then distributed to two heads: the original separation decoder for source reconstruction and a 2-layer transformer-based diarization head. The latter produces per-frame speaker probabilities, $\hat{y}_{\text{spk}}(t)$, using 8-head self-attention and position-wise feed-forward networks.

The model is optimized using a multi-task learning objective. The total loss $\mathcal{L}_{\text{total}}$ is defined as a weighted sum of the separation and diarization losses:
\begin{equation}
    \mathcal{L}_{\text{total}} = \mathcal{L}_{\text{SI-SNR}} + \lambda_{\text{diar}} \cdot \mathcal{L}_{\text{diarization}}
\end{equation}
where $\lambda_{\text{diar}} = 20.0$ serves as a balancing hyperparameter. We selected this value empirically after testing a range of candidates from $0.5$ to $20$. The diarization loss $\mathcal{L}_{\text{diar}}$ is computed using frame-level cross-entropy between the predicted speaker probabilities $\hat{y}_{\text{spk}}(t)$ and the ground truth labels $y_{\text{spk}}(t)$:
\begin{equation}
    \mathcal{L}_{\text{diar}} = \text{CrossEntropy}(\hat{y}_{\text{spk}}(t), y_{\text{spk}}(t))
\end{equation}
This shared representation learning is intended to facilitate knowledge transfer between the separation and diarization tasks while providing an implicit regularization effect. To investigate this interaction, we employed a phased fine-tuning schedule: the separation backbone was initially frozen for 10 epochs to allow the diarization head to stabilize on the pre-trained features. Subsequently, the entire module was unfrozen for 5 additional epochs.

\section{Experimental Setup}
\label{experiments}

\subsection{Datasets}
\label{datasets}
We evaluate our methods on both synthetic and organic multi-talker datasets to ensure robustness across varying acoustic conditions.

\textbf{Synthetic Mixtures:} We utilize Libri2Mix \cite{cosentino2020librimix}, consisting of fully overlapping speech with (\textit{both}) and without (\textit{clean}) WHAM noise \cite{Wichern2019WHAM}. We use the 3,000-sample test sets for evaluation and the \textit{both-train} subset for fine-tuning the exploratory diarization head. Additionally, we test on LibriSpeechMix \cite{kanda2020serialized}, which features partially overlapping clean speech, providing a more complex temporal structure than the fully overlapped LibriMix.

\textbf{Real-World Data:} We employ the AMI Meeting Corpus, focusing on the Single Distant Microphone (SDM) and Individual Headset Mix (IHM) scenarios. Audio is partitioned into ``utterance groups" following \cite{Kanda2021}. To maintain consistency with our separation backbones, we retain only two-speaker groups, resulting in a test set of 659 segments.

\subsection{Evaluation Metric}
\label{metric}

To evaluate system performance, we employ the Concatenated Permutation Word Error Rate ($cpWER$), the standard metric for multi-talker ASR. $cpWER$ addresses speaker order ambiguity by calculating the $WER$ for all possible speaker permutations $\pi \in \Pi$ and selecting the minimum.

\begin{equation}
    \text{cpWER} = \min_{\pi \in \Pi} \left( \frac{\sum_{i=1}^{n} (S_i + D_i + I_i)}{\sum_{i=1}^{n} N_i} \right)
\end{equation}
where n is the number of references/speakers ($n=2$ in this study), $N_i$ is the word count of the $i$-th reference, and $S$, $D$, and $I$ represent the number of substitutions, deletions, and insertions, respectively. For brevity, we refer to $cpWER$ simply as WER throughout the remainder of this paper.

\subsection{Implementation Details}
\label{implementation_details}
To implement our testing pipeline, we utilized several state-of-the-art pretrained models. These models were chosen following an initial selection process that considered their prominence in the literature, the supplementary information they produce, and their performance in preliminary experiments.

\textbf{Speech Separation: }To isolate individual speech sources from the mixtures, we evaluated two state-of-the-art pretrained models, namely Mossformer2 \cite{Mossformer2024} and Sepformer \cite{sepformer2021}. Mossformer2 was trained on large datasets, including synthetic data and data with a high level of background noise, making it a reliable choice for achieving top-tier accuracy even in acoustically challenging environments. On the other hand, Sepformer is fast due to its parallelism capability and uses low memory, making it the perfect model for a low-resource development environment. We employed the version pretrained on Libri2Mix, namely sepformer-libri2mix.\footnote{https://huggingface.co/speechbrain/sepformer-libri2mix}

\textbf{Speech Recognition:} After separating the speech sources, we utilized Universal-2 \cite{assemblyai2024} to transcribe them. Universal-2 is a model designed to meet the requirements of large-scale, multilingual ASR. It was shown to be superior to other state-of-the-art models such as Whisper \cite{whisper2023} in various domains, such as code-switching, inference speed, hallucination rates, etc. Besides its performance, our choice for Universal-2 was also motivated by its ability to predict word-level timestamps, which are necessary for our proposed transcript correction algorithm to function.

\textbf{Diarization Verifier:} For our proposed pruning-based transcription method depicted in Section \ref{diarization_based_correction_method}, we utilized \textit{pyannote.audio 3.1} \cite{Bredin23pyannote} as the Diarizer. It is open source, widely used in the literature, and has been benchmarked on a large collection of datasets, including the AMI meeting corpus used in our experiments.

\section{Results and Discussions}
\label{sec:results_and_discussions}

\subsection{Multi-Talker ASR Results}
\label{subsec:mt_asr_results}

\begin{table*}[]
\centering
\caption{Comparison of $WER$ (\%) across datasets. \textcolor{green}{$\downarrow$} indicates a performance improvement, while \textcolor{red}{$\uparrow$} indicates a decline. The improvement/decline percentage for \textit{Joint Mossformer-Diar} $\rightarrow$ Universal-2 is relative to the Mossformer $\rightarrow$ Universal-2 baseline.}
\label{tab:comparison_of_cpWER_accross_datasets}
\resizebox{\textwidth}{!}{%
\begin{tabular}{lccccc}
\hline
\multirow{2}{*}{System} & \multicolumn{2}{c}{Libri2Mix} & LibriSpeechMix & \multicolumn{2}{c}{AMI} \\ \cline{2-6} 
 & test-clean & test-both & 2spk & SDM & IHM \\ \hline
\textit{Joint Mossformer-Diar} & & & & & \\
\quad $\hookrightarrow$ Universal-2 (experimental) & 3.52 {\scriptsize (\textcolor{green!70!black}{$\downarrow$} 0.28\%)} & 9.23 {\scriptsize (\textcolor{green!70!black}{$\downarrow$} 0.86\%)} & 6.10 {\scriptsize (\textcolor{red}{$\uparrow$} 18.91\%)} & 34.77 {\scriptsize (\textcolor{green!70!black}{$\downarrow$} 2.06\%)} & 28.94 {\scriptsize (\textcolor{red}{$\uparrow$} 20.78\%)} \\ \hline
\textit{Sepformer} & & & & & \\
\quad $\hookrightarrow$ Universal-2 (baseline) & 3.24 & 16.96 & 3.71 & 42.10 & 39.80 \\
\quad \quad $\hookrightarrow$ Spk-leakage fix (proposed) & \textbf{3.20} {\scriptsize (\textcolor{green!70!black}{$\downarrow$} 1.23\%)} & \textbf{16.90} {\scriptsize (\textcolor{green!70!black}{$\downarrow$} 0.35\%)} & \textbf{3.53} {\scriptsize (\textcolor{green!70!black}{$\downarrow$} 4.85\%)} & \textbf{39.66} {\scriptsize (\textcolor{green!70!black}{$\downarrow$} 5.80\%)} & \textbf{35.60} {\scriptsize (\textcolor{green!70!black}{$\downarrow$} 10.55\%)} \\ 
\textit{Mossformer} & & & & & \\
\quad $\hookrightarrow$ Universal-2 (baseline) & 3.53 & 9.31 & 5.13 & 35.50 & 23.96 \\
\quad \quad $\hookrightarrow$ Spk-leakage fix (proposed) & \textbf{3.44} {\scriptsize (\textcolor{green!70!black}{$\downarrow$} 2.55\%)} & \textbf{9.17} {\scriptsize (\textcolor{green!70!black}{$\downarrow$} 1.50\%)} & \textbf{4.64} {\scriptsize (\textcolor{green!70!black}{$\downarrow$} 9.55\%)} & \textbf{32.67} {\scriptsize (\textcolor{green!70!black}{$\downarrow$} 7.97\%)} & \textbf{22.26} {\scriptsize (\textcolor{green!70!black}{$\downarrow$} 7.10\%)} \\ \hline
\end{tabular}%
}
\end{table*}

The experimental results (Table \ref{tab:comparison_of_cpWER_accross_datasets}) demonstrate that the proposed speaker leakage fix consistently shows improved performance by reducing $WER$ across all test conditions. For both Sepformer and Mossformer backbones, the fix achieves its most significant gains on real-world meeting data, notably yielding a relative improvement of $10.55\%$ on the AMI IHM set for Sepformer and $7.97\%$ on the AMI SDM set for Mossformer.

A critical observation is the performance disparity of the \emph{Joint Mossformer-Diar}. While it shows minor improvements on the noisier Libri2Mix and AMI SDM datasets, it suffers significant degradation on LibriSpeechMix ($+18.91\%$) and AMI IHM ($+20.78\%$). Since the model was fine-tuned on Libri2Mix-both—which contains high levels of synthetic noise—the joint head appears to benefit from the more challenging, far-field acoustic conditions of SDM. However, it fails to generalize to the ``cleaner" signal profiles of LibriSpeechMix and IHM. This suggests the joint architecture is highly sensitive to the acoustic domain and SNR of the training data. In contrast, our proposed pruning algorithm remains robust to these varying channel conditions, providing a consistent improvement across both simulated and organic datasets.

\begin{table}[ht]
\centering
\caption{WER (\%) comparison for high-leakage samples (transcripts similarity $> 0.4$). Gray values ($N$) indicate sample counts; \textcolor{green!70!black}{$\downarrow$} denotes relative reduction compared to baseline.}
\label{tab:comparison_of_cpWER_accross_leakage_subsets}
\setlength{\tabcolsep}{2pt} 
\resizebox{\columnwidth}{!}{%
\begin{tabular}{clccc}
\hline
& \textbf{Dataset} & \textbf{Baseline} & \textbf{Leakage Fix} & \textbf{Rel. $\Delta$} \\ \hline
\multirow{5}{*}{\rotatebox[origin=c]{90}{\textbf{Sepformer}}} 
& L2M test-clean {\color{gray}\scriptsize ($N=64$)} & 16.88 & \textbf{14.33} & \textcolor{green!70!black}{$\downarrow$ 15.11\%} \\
& L2M test-both {\color{gray}\scriptsize ($N=65$)}  & 31.25 & \textbf{28.19} & \textcolor{green!70!black}{$\downarrow$ 9.79\%} \\
& LSM 2spk {\color{gray}\scriptsize ($N=84$)}       & 29.64 & \textbf{21.54} & \textcolor{green!70!black}{$\downarrow$ 27.33\%} \\
& AMI SDM {\color{gray}\scriptsize ($N=138$)}        & 71.53 & \textbf{57.54} & \textcolor{green!70!black}{$\downarrow$ 19.56\%} \\
& AMI IHM {\color{gray}\scriptsize ($N=175$)}        & 77.75 & \textbf{62.22} & \textcolor{green!70!black}{$\downarrow$ 19.97\%} \\ \hline
\multirow{5}{*}{\rotatebox[origin=c]{90}{\textbf{Mossformer}}} 
& L2M test-clean {\color{gray}\scriptsize ($N=79$)} & 18.82 & \textbf{14.74} & \textcolor{green!70!black}{$\downarrow$ 21.68\%} \\
& L2M test-both {\color{gray}\scriptsize ($N=88$)}  & 33.39 & \textbf{27.41} & \textcolor{green!70!black}{$\downarrow$ 17.91\%} \\
& LSM 2spk {\color{gray}\scriptsize ($N=120$)}      & 48.58 & \textbf{34.51} & \textcolor{green!70!black}{$\downarrow$ 28.96\%} \\
& AMI SDM {\color{gray}\scriptsize ($N=122$)}        & 76.12 & \textbf{57.98} & \textcolor{green!70!black}{$\downarrow$ 23.83\%} \\
& AMI IHM {\color{gray}\scriptsize ($N=76$)}         & 65.58 & \textbf{46.39} & \textcolor{green!70!black}{$\downarrow$ 29.26\%} \\ \hline
\end{tabular}%
}
\end{table}

To further evaluate the precision of the proposed transcript correction method, a targeted analysis was performed on the subset of samples identified as containing significant speaker leakage, defined by a separated source transcript similarity greater than $0.4$. The results in Table \ref{tab:comparison_of_cpWER_accross_leakage_subsets} provide several key insights:

\textbf{Targeted Error Mitigation:} The high baseline—e.g., $WER$ $> 70\%$ in AMI-SDM—confirms that speaker leakage is a primary driver of ASR failure. The proposed fix achieves dramatic relative reductions, peaking at $29.26\%$ for Mossformer, demonstrating high precision in correcting the intended failure mode.

\textbf{Architectural Robustness:} Despite variations in the number of identified leaked samples ($N$) across backbones—e.g., $N=120$ for Mossformer vs. $N=84$ for Sepformer in LibriSpeechMix—the fix provides consistent, large-scale improvements, proving its versatility as a post-processing enhancement.

\textbf{Real-World Generalization:} The most significant absolute gains occur in the AMI corpus, where $WER$ drops from $65.58\%$ to $46.39\%$ (Mossformer IHM). This highlights the module's efficacy in organic meeting environments characterized by complex, partial overlaps compared to fully overlapping simulated data.

\subsection{Ablation Study}
\label{ablation_study}

We conducted an ablation study to investigate the relative contributions of the modalities to the performance of the proposed speaker leakage fix. As shown in Table \ref{tab:ablation_study}, the \textit{Text Only} configuration, which prunes words based solely on their presence in the parallel transcript ($C_{\text{lex}}$), consistently degrades performance relative to the baseline. Without the signal-level grounding of $C_{\text{ac}}$ or the time-sensitive constraint of $C_{\text{temp}}$, this modality suffers from extreme \textit{over-correction}, erroneously removing words that are common to both speakers but correctly transcribed.

Conversely, the \textit{Acoustic Only} variant, which relies only on $C_{\text{ac}}$, provides substantial WER reductions, confirming that signal-level diarization windows are the primary driver of leakage detection. However, the most robust results are consistently achieved by the full \textit{Text + Acoustic} system. By requiring the tripartite consensus ($C_{\text{ac}} \land C_{\text{lex}} \land C_{\text{temp}}$), the model effectively uses lexical and temporal alignment as safety gates. This ensures that only high-confidence leakage artifacts are pruned, while unique spoken content that may overlap with acoustic leakage windows is preserved. This validates the necessity of a multimodal approach where acoustic features provide detection and lexical cues provide essential grounding.

\begin{table}[ht]
\centering
\caption{Ablation study of the Speaker Leakage Fix on high-leakage subsets (transcripts similarity $> 0.4$). We compare the baseline against fixes using text-only, acoustic-only, and combined text + acoustic modalities. All values are $WER$ (\%).}
\label{tab:ablation_study}
\setlength{\tabcolsep}{2.5pt} 
\resizebox{\columnwidth}{!}{%
\begin{tabular}{clcccc}
\hline
& & \multicolumn{1}{c}{} & \multicolumn{3}{c}{\textbf{Modality}} \\ \cline{4-6}
& \textbf{Dataset} & \textbf{Baseline} & \textbf{Text} & \textbf{Acoustic} & \textbf{All} \\ \hline
\multirow{5}{*}{\rotatebox[origin=c]{90}{\textbf{Sepformer}}} 
& Libri2Mix test-clean & 16.88 & 31.54 & 18.46 & \textbf{14.33} \\
& Libri2Mix test-both  & 31.25 & 44.19 & 38.41 & \textbf{28.19} \\
& LibriSpeechMix 2spk  & 29.64 & 41.58 & 23.35 & \textbf{21.54} \\
& AMI SDM              & 71.53 & 78.17 & 58.93 & \textbf{57.54} \\
& AMI IHM              & 77.75 & 82.87 & \textbf{59.84} & 62.22 \\ \hline
\multirow{5}{*}{\rotatebox[origin=c]{90}{\textbf{Mossformer}}} 
& Libri2Mix test-clean & 18.82 & 32.53 & 21.19 & \textbf{14.74} \\
& Libri2Mix test-both  & 33.39 & 46.72 & 37.65 & \textbf{27.41} \\
& LibriSpeechMix 2spk  & 48.58 & 57.20 & 36.58 & \textbf{34.51} \\
& AMI SDM              & 76.12 & 83.54 & 60.13 & \textbf{57.98} \\
& AMI IHM              & 65.58 & 76.54 & 46.45 & \textbf{46.39} \\ \hline
\end{tabular}%
}
\end{table}

\section{Conclusions}
\label{conclusions}
In this paper, we addressed the persistent challenge of speaker leakage in cascaded multi-talker ASR systems. Unlike traditional post-processing methods that rely on lexical re-labeling—which can become inherently ambiguous in high-leakage scenarios—we introduced a novel pruning-based paradigm. By utilizing a pre-trained diarization model as a multimodal verifier, our algorithm effectively identifies and removes spurious speech artifacts through a tripartite consensus of acoustic, lexical, and temporal evidence.

Our experimental results across synthetic and real-world datasets demonstrate the robustness of our proposed algorithm. It achieved significant WER reductions, with relative improvements of up to 29\% on high-leakage subsets of the AMI Meeting corpus. It proved to be a versatile and computationally efficient enhancement that generalizes well to organic meeting environments. 


\clearpage
\newpage

\section{Generative AI Use Disclosure}
The use of generative AI was limited to editing and polishing the manuscript.

\bibliographystyle{IEEEtran}
\bibliography{mybib}

@InProceedings{whisper2023,
	title = 	 {Robust Speech Recognition via Large-Scale Weak Supervision},
	author =       {Radford, Alec and Kim, Jong Wook and Xu, Tao and Brockman, Greg and Mcleavey, Christine and Sutskever, Ilya},
	booktitle = 	 {Proceedings of the 40th International Conference on Machine Learning},
	pages = 	 {28492--28518},
	year = 	 {2023},
	editor = 	 {Krause, Andreas and Brunskill, Emma and Cho, Kyunghyun and Engelhardt, Barbara and Sabato, Sivan and Scarlett, Jonathan},
	volume = 	 {202},
	series = 	 {Proceedings of Machine Learning Research},
	month = 	 {23--29 Jul},
	publisher =    {PMLR},
	url = 	 {https://proceedings.mlr.press/v202/radford23a.html}
}

@article{assemblyai2024,
	author = {Ramirez, Francis McCann and Chkhetiani, Luka and Ehrenberg, Andrew and McHardy, Robert and Botros, Rami and Khare, Yash and Vanzo, Andrea and Peyash, Taufiquzzaman and Oexle, Gabriel and Liang, Michael and Sklyar, Ilya and Fakhan, Enver and Etefy, Ahmed and McCrystal, Daniel and Flamini, Sam and Donato, Domenic and Yoshioka, Takuya},
	ee = {https://doi.org/10.48550/arXiv.2404.09841},
	journal = {CoRR},
	title = {Anatomy of Industrial Scale Multilingual ASR.},
	url = {http://dblp.uni-trier.de/db/journals/corr/corr2404.html#abs-2404-09841},
	volume = {abs/2404.09841},
	year = 2024
}

@inproceedings{Kanda2021,
	title={Large-Scale Pre-Training of End-to-End Multi-Talker ASR for Meeting Transcription with Single Distant Microphone},
	author={Naoyuki Kanda and Guoli Ye and Yu Wu and Yashesh Gaur and Xiaofei Wang and Zhong Meng and Zhuo Chen and Takuya Yoshioka},
	booktitle={Proc. Interspeech},
	year={2021},
	url={https://api.semanticscholar.org/CorpusID:232427909}
}

@INPROCEEDINGS{Mossformer2024,
  author={Zhao, Shengkui and Ma, Yukun and Ni, Chongjia and Zhang, Chong and Wang, Hao and Nguyen, Trung Hieu and Zhou, Kun and Yip, Jia Qi and Ng, Dianwen and Ma, Bin},
  booktitle={Proc. IEEE ICASSP}, 
  title={MossFormer2: Combining Transformer and RNN-Free Recurrent Network for Enhanced Time-Domain Monaural Speech Separation}, 
  year={2024},
  volume={},
  number={},
  pages={10356-10360},
  doi={10.1109/ICASSP48485.2024.10445985}}

@inproceedings{Wichern2019WHAM,
    title     = {WHAM!: Extending Speech Separation to Noisy Environments},
    author    = {Wichern, Gordon and Antognini, Joe and Flynn, Michael and Zhu,
                 Licheng Richard and McQuinn, Emmett and Crow,
                 Dwight and Manilow, Ethan and Le Roux, Jonathan},
    booktitle = {Proc. Interspeech},
    year      = {2019},
    month     = sep
}

@inproceedings{Meng2024whisper,
	title={{Empowering Whisper as a Joint Multi-Talker and Target-Talker Speech Recognition System}},
	author={Meng, Lingwei and Kang, Jiawen and Wang, Yuejiao and Jin, Zengrui and Wu, Xixin and Liu, Xunying and Meng, Helen},
	booktitle={Proc. INTERSPEECH},
	year={2024}
}

@INPROCEEDINGS{Thai-Binh2024,
	author={Nguyen, Thai-Binh and Waibel, Alexander},
	booktitle={Proc. IEEE ICASSP}, 
	title={Synthetic Conversations Improve Multi-Talker ASR}, 
	year={2024},
	volume={},
	number={},
	pages={10461-10465},
	doi={10.1109/ICASSP48485.2024.10446589}}

@inproceedings{kanda2020serialized,
	title={Serialized Output Training for End-to-End Overlapped Speech Recognition},
	author={Kanda, Naoyuki and Gaur, Yashesh and Wang, Xiaofei and Meng, Zhong and Yoshioka, Takuya},
	booktitle={Proc. Interspeech},
	pages={2797--2801},
	year={2020}
}

@misc{cosentino2020librimix,
	title={LibriMix: An Open-Source Dataset for Generalizable Speech Separation},
	author={Joris Cosentino and Manuel Pariente and Samuele Cornell and Antoine Deleforge and Emmanuel Vincent},
	year={2020},
	eprint={2005.11262},
	archivePrefix={arXiv},
	primaryClass={eess.AS}
}

@INPROCEEDINGS{sepformer2021,
	author={Subakan, Cem and Ravanelli, Mirco and Cornell, Samuele and Bronzi, Mirko and Zhong, Jianyuan},
	booktitle={Proc. IEEE ICASSP}, 
	title={Attention Is All You Need In Speech Separation}, 
	year={2021},
	volume={},
	number={},
	pages={21-25},
	doi={10.1109/ICASSP39728.2021.9413901}}

@INPROCEEDINGS{Yufeng2025,
	author={Yang, Yufeng and Taherian, Hassan and Kalkhorani, Vahid Ahmadi and Wang, DeLiang},
	booktitle={ICASSP}, 
	title={Elevating Robust ASR By Decoupling Multi-Channel Speaker Separation and Speech Recognition}, 
	year={2025},
	volume={},
	number={},
	pages={1-5},
	doi={10.1109/ICASSP49660.2025.10888074}}

@inproceedings{Paturi2023LexicalSE,
  title={Lexical Speaker Error Correction: Leveraging Language Models for Speaker Diarization Error Correction},
  author={Rohit Paturi and Sundararajan Srinivasan and Xiang Li},
  booktitle={Interspeech},
  year={2023},
  url={https://api.semanticscholar.org/CorpusID:259171762}
}

@article{paturi2024aglsec,
  title={Ag-lsec: Audio grounded lexical speaker error correction},
  author={Paturi, Rohit and Li, Xiang and Srinivasan, Sundararajan},
  journal={arXiv preprint arXiv:2406.17266},
  year={2024}
}

@article{He2025InteractiveRS,
  title={Interactive Real-Time Speaker Diarization Correction with Human Feedback},
  author={Xinlu He and Yiwen Guan and Badrivishal Paurana and Zilin Dai and Jacob Whitehill},
  journal={ArXiv},
  year={2025},
  volume={abs/2509.18377},
  url={https://api.semanticscholar.org/CorpusID:281496027}
}

@INPROCEEDINGS{Turn-to-Diarize2022,
  author={Xia, Wei and Lu, Han and Wang, Quan and Tripathi, Anshuman and Huang, Yiling and Moreno, Ignacio Lopez and Sak, Hasim},
  booktitle={ICASSP 2022 - 2022 IEEE International Conference on Acoustics, Speech and Signal Processing (ICASSP)}, 
  title={Turn-to-Diarize: Online Speaker Diarization Constrained by Transformer Transducer Speaker Turn Detection}, 
  year={2022},
  volume={},
  number={},
  pages={8077-8081},
  doi={10.1109/ICASSP43922.2022.9746531}}

@article{Adedeji2024TheSoundOfHealcare,
  title={The Sound of Healthcare: Improving Medical Transcription ASR Accuracy with Large Language Models},
  author={Ayo Adedeji and Sarita Joshi and Brendan Doohan},
  journal={ArXiv},
  year={2024},
  volume={abs/2402.07658},
  url={https://api.semanticscholar.org/CorpusID:267627126}
}

@article{Park2023EnhancingSD,
  title={Enhancing Speaker Diarization with Large Language Models: A Contextual Beam Search Approach},
  author={Tae Jin Park and Kunal Dhawan and Nithin Rao Koluguri and Jagadeesh Balam},
  journal={ICASSP 2024 - 2024 IEEE International Conference on Acoustics, Speech and Signal Processing (ICASSP)},
  year={2023},
  pages={10861-10865},
  url={https://api.semanticscholar.org/CorpusID:261681807}
}

@article{Wang2024DiarizationLMSD,
  title={DiarizationLM: Speaker Diarization Post-Processing with Large Language Models},
  author={Quan Wang and Yiling Huang and Guanlong Zhao and Evan Clark and Wei Xia and Hank Liao},
  journal={ArXiv},
  year={2024},
  volume={abs/2401.03506},
  url={https://api.semanticscholar.org/CorpusID:266844688}
}

@article{Kumar2025SEALSE,
  title={SEAL: Speaker Error Correction using Acoustic-conditioned Large Language Models},
  author={Anurag Kumar and Rohit Paturi and Amber Afshan and Sundararajan Srinivasan},
  journal={ICASSP 2025 - 2025 IEEE International Conference on Acoustics, Speech and Signal Processing (ICASSP)},
  year={2025},
  pages={1-5},
  url={https://api.semanticscholar.org/CorpusID:275544302}
}

@INPROCEEDINGS{continuous_ss_w_conformer2021,
  author={Chen, Sanyuan and Wu, Yu and Chen, Zhuo and Wu, Jian and Li, Jinyu and Yoshioka, Takuya and Wang, Chengyi and Liu, Shujie and Zhou, Ming},
  booktitle={ICASSP 2021 - 2021 IEEE International Conference on Acoustics, Speech and Signal Processing (ICASSP)}, 
  title={Continuous Speech Separation with Conformer}, 
  year={2021},
  volume={},
  number={},
  pages={5749-5753},
  doi={10.1109/ICASSP39728.2021.9413423}}

@article{Paturi2021DirectedSS,
  title={Directed Speech Separation for Automatic Speech Recognition of Long Form Conversational Speech},
  author={Rohit Paturi and Sundararajan Srinivasan and Katrin Kirchhoff},
  journal={ArXiv},
  year={2021},
  volume={abs/2112.05863},
  url={https://api.semanticscholar.org/CorpusID:245123727}
}

@article{Wang2025SpeakerTV,
  title={Speaker Targeting via Self-Speaker Adaptation for Multi-talker ASR},
  author={Weiqing Wang and Tae Jin Park and Ivan Medennikov and Jinhan Wang and Kunal Dhawan and He Huang and Nithin Rao Koluguri and Jagadeesh Balam and Boris Ginsburg},
  journal={ArXiv},
  year={2025},
  volume={abs/2506.22646},
  url={https://api.semanticscholar.org/CorpusID:280016287}
}

@article{shi2025serializedoutputprompting,
  title={Serialized Output Prompting for Large Language Model-based Multi-Talker Speech Recognition},
  author={Shi, Hao and Fujita, Yusuke and Mizumoto, Tomoya and Liu, Lianbo and Kojima, Atsushi and Sudo, Yui},
  journal={arXiv preprint arXiv:2509.04488},
  year={2025}
}

@inproceedings{Bredin23pyannote,
  author={Hervé Bredin},
  title={{pyannote.audio 2.1 speaker diarization pipeline: principle, benchmark, and recipe}},
  year=2023,
  booktitle={Proc. INTERSPEECH 2023},
}

@inproceedings{vieting2025error,
  title={Error Analysis in a Modular Meeting Transcription System},
  author={Vieting, Peter and Berger, Simon and von Neumann, Thilo and Boeddeker, Christoph and Schl{\"u}ter, Ralf and Haeb-Umbach, Reinhold},
  booktitle={Speech Communication; 16th ITG Conference},
  pages={1--5},
  year={2025},
  organization={VDE}
}

@article{black2004ratcliff,
  title={Ratcliff/Obershelp pattern recognition},
  author={Black, Paul E},
  journal={Dictionary of algorithms and data structures},
  volume={17},
  year={2004},
  publisher={National Institute of Standards and Technology Gaithersburg, MD, USA}
}

\end{document}